\documentclass[aps,prl,twocolumn,showpacs,superscriptaddress,footinbib]{revtex4-1}
\usepackage{graphicx}% Include figure files
\usepackage{dcolumn}% Align table columns on decimal point
\usepackage{bm}% bold math
\usepackage{enumerate}
\usepackage{amsmath}

\usepackage[normalem]{ulem}  % \sout{old text} for strikeout
\begin{document}

%\preprint{APS/123-QED}

\title{A simple explanation of the absence of the spherical nuclei with the hidden-color states}

\author{Tao Wang}
\email{suiyueqiaoqiao@163.com}
\affiliation{College of Physics, Tonghua Normal University, Tonghua 134000, People's Republic of China}

\date{\today}

\begin{abstract}
\textbf{Abstract:} Inspired by the EMC effect, the Cd puzzle and the SU3-IBM, a hypothesis can be given for a nucleus, that only the nucleus itself is a trivial (0,0) representation of the SU(3) group, which leads to the simple conclusion that spherical nucleus does not exist and the spherical mean field is not allowed. The key conclusion is that the whole nucleus should have many hidden-color states, and the quark-gluon degrees of freedom are required even at the low-energy excitation of the nucleus.

\end{abstract}

\maketitle

\section{1. Introduction}

The past decade has seen some perspective shifts in the field of nuclear physics. An important one is the discovery of the Cd puzzle \cite{Wood11}, which makes the traditional perspective on the collective mode of phonon excitation of the nucleus with spherical shape to be questioned and refuted \cite{Heyde16,Garrett18}. On the basis of these related experiments \cite{Garrett08,Garrett10,Garrett12,Batchelder12,Garrett19,Garrett20}, I proposed that this is a new mode of collective excitation, and found that the extended interacting boson model \cite{Iachello87,Casten10,Wang08} with SU(3) higher-order interactions (SU3-IBM) can be used to explain this new mode \cite{Wang22,WangPt,Wang24}. These findings lead us to question the traditional perspective on the emergence of the collective behaviors in the nucleus. Since the Cd puzzle conflicts with previous ideas and the SU3-IBM is very different from previous nuclear structure theories, it can be inferred that some \emph{global} effects in the nucleus may be neglected in previous studies and the emergence of the collectivity in the nucleus may have an implicit origin.

The prominent SU(3) symmetry appearing in the SU3-IBM provides a unique clue, for which one will doubt whether this will be related to the color SU(3) symmetry in the quantum chromodynamics \cite{Drischler21,Gross23,Achenbach24}. In existing studies, especially after the discovery of the EMC effect \cite{Arneodo94,Thoma95,Weise00,Hen17}, some researchers began to understand some phenomena in the nucleus from the quark-gluon degree of freedom, and especially the nuclear forces at short distances may require the quark-gluon interactions to explain their repulsion \cite{Bakker14,Abbas17,Sargsian03,Sargsian14,Clot19}. But no one believed that the low-energy excitation of the nucleus is related to the color degree of freedom. In this Letter, I establish such a connection for the first time. If the hidden-color states exist \cite{Harvey81,Harvey812,Ji86,Abbas86,Myhrer88,Ping11,Miller14,Miller19,Miller20,Entem20}, the results of these experiments for the absence of the spherical nuclei can have a very simple explanation. While some researchers have discussed the hidden-color state in the nucleus with fewer nucleons, no one was sure that the whole nucleus itself should have many hidden-color states. This implies that the shape deformation of the nucleus may arise from the quark-gluon short-distance interactions between the nucleons in the nucleus. Although this conclusion is somewhat different from previous ideas, it is actually reasonable.

\section{2. The Cd puzzle and the SU3-IBM}

In the traditional wisdom, nuclei with magic numbers are spherical \cite{Otsuka22,Wood22}, because the magic numbers can be well explained by the spherical central three-dimensional harmonic oscillator potential with strong spin-orbital coupling \cite{Mayer55}. This mean-field view, or the shell model, forms the basis of the nuclear structure \cite{Wood22}, which ``is the fundamental starting point for all effects to provide a theory of nuclei'' \cite{Hen17}. When the valence nucleons increase (the number is still small), the spherical surface is still kept, and the residual two-body interaction is expected to play a dominant role, which leads to the phonon vibrational excitation for the even-even nucleus \cite{Heyde16,Garrett18,Otsuka,Bohr75}.

The Cd nuclei have long been considered as typical example with spherical vibrational excitation \cite{Garrett18}. The reason is that the low-lying levels of these nuclei look very similar to the phonon excitation spectra. However the shape coexistence exists in the Cd nuclei \cite{Cohen60,Meyer77}, which hinders the understanding of the normal states of the Cd isotopes. It was long recognized that the electromagnetic transitions in the normal states are different from the ones in the spherical phonon mode. Since researchers in the field of nuclear structure believed that the Cd nuclei must be spherical, this deviation is suggested to result from the strong coupling between the normal states and intruder states \cite{Heyde82,Kern93,Jolie93}.

No one believed that there may be other possibility until the Cd puzzle experiments denied this idea \cite{Garrett08,Garrett10,Garrett12,Batchelder12,Garrett19,Garrett20}. The coupling between the normal states and the intruder states is in fact weak \cite{Garrett12}, and the spherical phonon excitation of the Cd nuclei is refuted \cite{Heyde16,Garrett18,Wood11}. Just as Garret, Green and Wood said: ``it is possible that the Cd isotopes may not represent vibrational systems and that the essential physics of their motion has been missing'' \cite{Garrett08}. I found that the normal states of the Cd nuclei present a new spherical-like $\gamma$-soft collective mode \cite{Wang22,WangPt,Wang24}. This means that the origin of the collective nature in the nucleus requires an alternative explanation. Just as Heyde and Wood said: ``The emerging picture of nuclear shapes is that quadrupole deformation is fundamental to achieving a unified view of nuclear structure.'' and ``a shift in perspective is needed: sphericity is a special case of deformation.'' \cite{Heyde16}

The Cd puzzle challenges the conventional view of nuclear structure, because these experimental results are more than expected. Inspired by the Cd puzzle, I proposed the SU3-IBM \cite{Wang22}, just satisfying the idea of Heyde and Wood. In this model, only the U(5) limit and the SU(3) limit are included. It suggests that the SU(3) symmetry dominates the onset of the quadrupole deformation \cite{wang23,zhou23}. Thus not only the quadrupole deformation but also the SU(3) symmetry is important for the unified description of nuclear structure. The Cd puzzle really presents a new spherical-like deformation \cite{Wang22,WangPt,Wang24}.

In this new spherical-like $\gamma$-soft rotation, it looks very similar to the phonon excitation spectrum of the spherical shape except that the $0_{3}^{+}$ state is repelled to a higher energy level \cite{Wang22,WangPt}. Thus at the previous three-phonon level, the $0_{3}^{+}$ state does not exist at all \cite{Wang24}. The B(E2) values among the normal states can be well explained in this model. In particular, it was recently found that the anomalous evolution trend of the electric quadrupole moment of the $2_{1}^{+}$ state of the $^{108-116}$Cd can be explained by this model, which implies that the SU3-IBM is the valuable attempt to resolve the Cd puzzle \cite{Wang24}.

Therefore, the current situation is that both experiment and theory are beginning to support the conclusion that the spherical nucleus does not exist. But it also raises new problems. There are two important conclusions in the SU3-IBM: (1) higher-order interactions are important and the third-order interaction can be even dominant \cite{Wang22,wang23,Wang}, (2) the SU(3) symmetry governs all the interactions except for the pairing force (the $d$ boson number operator). Thus the idea of the mean field requires some corrections that the two-body interaction may not necessarily dominate the residual ones. And an important question is, where does this SU(3) symmetry in the SU3-IBM come from?

In previous IBM, the SU(3) symmetry is only one reduction limit of the SU(6) group \cite{Iachello87}, so it is an accidental product of the shape symmetry spontaneous breaking of the nucleus (from the spherical shape to the prolate shape) \cite{Cseh19}. Its appearance does not require an explanation. But in the SU3-IBM, this situation becomes different, because all the quadrupole shapes are described by the SU(3) symmetry, and nearly the whole Hamiltonian has the SU(3) symmetry except for the pairing one, so a deeper understanding is needed.

Obviously, the success of the SU3-IBM proves that the key to the Cd puzzle, the fact that the spherical nucleus does not exist, is related to the SU(3) symmetry. This result is very important for a further deeper understanding of the origin of the Cd puzzle. In the SU(3) representation, (0,0) presents the spherical shape. Thus the Cd puzzle implies that the shape of the nucleus can not be the (0,0) representation of the SU(3) group. In this Letter, a simple but strict explanation on this impossibility is given.

The Cd puzzle is the low-energy phenomenon of the nucleus, which is related to the collective or \emph{global} effect of the nucleons in the nucleus. It is hard to imagine that this phenomenon may be related to the high-energy behavior. However, in the nucleus, especially for the low-energy excitation, the only one with the exact SU(3) symmetry is the color symmetry in the quantum chromodynamics \cite{Drischler21,Gross23,Achenbach24}, so the two must be linked together. Although somewhat beyond expectations, as the SU3-IBM was proposed to resolve the Cd puzzle, this is the \emph{unique} possibility. Here some unexpected relations are found. The EMC effect \cite{Arneodo94,Thoma95,Weise00,Hen17} means that quarks and gluons still play an important role in the nucleus. It is found that the color degrees of freedom is also required in the low-energy excitation of the nucleus.

\section{3. The EMC effect and the hidden-color state}

A Nucleus is made of protons and neutrons, which are collectively called nucleons. The nucleons are also composite particles, consisting of valence quarks, sea quarks, and gluons \cite{Drischler21,Gross23,Achenbach24}. Considering that the binding energy of the nucleus is less than one percent of the nuclear energy and the quark-gluon confinement in the free nucleons, many researchers would think that the nucleons in the nucleus are the same as the free nucleons, or treat them as point-like particles. However, the EMC effect found in 1983 rejected this perspective \cite{Arneodo94,Thoma95,Weise00,Hen17}.

The EMC effect illustrates that the result of deep inelastic scattering is related to the target of the used nucleus, and the effect increases as the mass number of the nucleus increases \cite{Arneodo94,Thoma95,Weise00,Hen17}. This means that the quark-gluon distribution of the nucleons in the nucleus is not the same as the ones in the free nucleons, which is related to the whole nucleus. It thus provides an example that the large-scale nucleus can affect the quark-gluon distribution in the small-scale nucleons. Just as Miller said: ``At least three length scales (the radius of the nucleus, the average inter-nucleon separation distance, and the size of the nucleon) are important in gaining a complete understanding of the physics of nuclei'' and ``the three scales are closely related, so that a narrow focus on any given specific range of scales
may prevent an understanding of the fundamental origins of nuclear properties'' \cite{Miller20}. Recently, researchers have generally agreed that this phenomenon is related to the medium modification of the nucleus, fundamentally, to the color confinement \cite{Hen17}.

Therefore in the nucleus, except for the nucleons and various instantaneous mesons, quarks and gluons need to be considered. There are two possibilities for the existence of this quark-gluon degree of freedom in nuclei. One is that although the nucleons in the nucleus change, the quarks and gluons are still bound in the nucleons. Another is, in the nucleus, quarks and gluons are not only bound within the nucleons, but also occasionally appear between the nucleons. For the latter, the nucleons in the nucleus may not be color-neutral, and have color. Of course, due to the color confinement, these colored nucleons must form a color-neutral whole entity. Since these quark-gluon interactions can only occur within the short distances between the nucleons, the united state with many colored nucleons in the nucleus is called the hidden-color state \cite{Abbas17}. Thus some researchers guess that ``multi-quark hidden-color components exist in ordinary nucleus'' \cite{Bakker14,Close82,Gabathuler85}.

Researchers have also studied the effect of the quarks and gluons in the nucleus \cite{Bakker14,Abbas17,Sargsian03,Sargsian14,Clot19}, and the hidden-color state \cite{Bakker14,Harvey81,Harvey812,Ji86,Abbas86,Myhrer88,Ping11,Miller14,Entem20,Miller19,Miller20}. Just as Clo\"{e}t \emph{et al.} said:`` it is unlikely that the nucleon-meson based approaches can remain valid or contain the correct degrees of freedom for all processes at all energy scales. Clearly identifying these scales and processes is key to exposing the role of quarks and gluons in nuclei and thereby developing an understanding of how nuclei emerge within QCD.'' \cite{Clot19} The quark-gluon degree of freedom is needed in the discussion of nuclei. The key is to find clear or revelatory evidence. It's hard to imagine that quarks and gluons are only confined to the interior of the nucleons and mesons while the nucleus has a larger volume. In this Letter I find the relationship between the absence of a spherical nucleus and the hidden-color states of the entire nucleus. For the existence of the Cd puzzle, the multi-quark hidden-color states should exist in ordinary nucleus.

Since the quark-gluon interactions can be only found within the short distances between the nucleons to produce a strong repulsive force that prevents the nucleons in the nucleus from collapsing into a much smaller sphere, that is, to maintain the stability of the nucleus \cite{Bakker14,Abbas17,Sargsian14}, this color force as the cause of the shape of the nucleus can be understandable. Without this effect, the shape of the nucleus would be meaningless.

\section{4. Explanation}

Quarks and gluons have color degrees of freedom which has SU(3) symmetry. The color confinement is a conjecture that has not been absolutely proved by theory. It means that composite particles made up of quarks and gluons, such as hadrons or nuclei, are color-neutral, or the (0,0) representations of the SU(3) group. Here I focus on the nuclei. Hydrogen nucleus is the proton, so it can be expressed by the (0,0) representation of the SU(3) group. The same is true for the neutron, and also true when extended to nuclei with larger mass number $A$. The whole composite of the quarks and gluons must be color-neutral. The colour confinement should be a \emph{global} phenomenon \cite{Abbas17}, which is the reason for the existence of the hidden-color state.

The EMC effect means that the nucleons in the nucleus are different from the free nucleons, which are the true embodiment of the color confinement. Although the specific physical mechanism is unknown, if the color confinement in the free nucleons is broken when bound in the nucleus, it is clear that these nucleons in the nucleus cannot be represented by the (0,0) representation of the SU(3) group. As long as the color degree of freedom appears between the nucleons in the nucleus, this conclusion is inevitable. It should also be emphasized that this color effect in nuclear medium becomes more pronounced as the number of nucleons increases.

The quark-gluon interactions are very necessary when discussing the short distance repulsions between the nucleons \cite{Bakker14,Abbas17,Sargsian03,Sargsian14,Clot19}. The hidden-color states has been also discussed in \cite{Bakker14,Harvey81,Harvey812,Ji86,Abbas86,Myhrer88,Ping11,Miller14,Entem20,Miller19,Miller20}. This conclusion is obvious.

Thus a hypothesis can be given for a nucleus, that only the nucleus itself is a trivial (0,0) representation of the SU(3) group. This means that the nucleons in the nucleus, and their combinations, as long as they are not the nucleus, cannot be the (0,0) representation of the SU(3) symmetry, no matter how complex the specific interactions between them are. Because if some of the nucleons are combined into the (0,0) representation, implying that this is a nucleus, it means that the original nucleus has been fission.

This assumption certainly needs the theoretical confirmation (color confinement), but here we will see that it directly gives the conclusion that a spherical nucleus does not exist. It has now been confirmed that the nuclear force can be divided into three parts: the mean field, short-range strong correlation, and long-range correlation \cite{Hen17}. The mean field presents the single particle effect, which means that the nucleons in the nucleus have some of the same features. The short-range strong correlation presents the correlation between only the two nucleons, and only the long-range correlation contains the collective or \emph{global} effect among all the nucleons in the nucleus.

Recent experiments have also demonstrated that the EMC effects are associated with the short-range strong correlations, but the more comprehensive correlations with the three parts remains unclear. It is already known that the EMC effects can be related to the corrections of the mean-field nucleons and short-range strongly correlated nucleons, but the exact mechanism is still unclear \cite{Hen17,Hen11,Hen12}. Thus some researchers will doubt whether such medium correction will be related with the low-energy nuclear structure. Recently Machleidt and Sammarruca said:`` it is a good question to ask whether medium modifications of nuclear forces show up in a noticeable way and/or are even needed for quantitative nuclear structure predictions." and asked: ``Are the energies typically involved in conventional nuclear structure physics low enough to treat nucleons as structure-less objects?" Here I give the answer, even at the low-lying excitation, it is not possible. The Cd puzzle means the nucleon modifications and the color degree freedom is also needed if the assumption is correct.

This explanation is very simple. If the representation of the nucleus itself is the (0,0) representation of the SU(3) group, then no part of it can be the (0,0) representation, which means that the mean field cannot be the (0,0) representation, and the long-range correlation is not the (0,0) representation too. Although this conclusion is simple, it is strict.

This requires a further detailed explanation. Since the representation of the final nucleus is (0,0), the product of the representation of all nucleons is (0,0), which has many possibilities if the color degree of freedom is considered. All possibilities need to be considered to understand the nature of the nucleus. Each possibility can be expressed as
\begin{equation}
(\lambda_{1},\mu_{1})\otimes (\lambda_{2},\mu_{2}) \otimes \cdot\cdot\cdot \otimes(\lambda_{A},\mu_{A}) =(0,0),
\end{equation}
here $A$ is the nucleon number, and $(\lambda_{k},\mu_{k})$ is the SU(3) representation of the kth nucleon, $1\leq k \leq A$. This is actually the expression of the color confinement in the nucleus. What is the $(\lambda_{k},\mu_{k})$ is still unknown, which needs further determined.

In most previous studies, the color degree of freedom is not considered, that is, the nucleons in the nucleus are color-neutral, then $(\lambda_{k},\mu_{k})$  in equation (1) are all (0,0). However if this equation (1) is a hidden-color state, they can not all be (0,0). Here the case that each $(\lambda_{k},\mu_{k})$ is not (0,0) is considered, which is the focus of this Letter. It should be stressed here that the intermediate coupling result can not appear as the (0,0) representation too.  When $A$ increases many coupling possibilities can be obtained for the hidden-color state, so there are in fact many hidden-color states with the mass number $A$ for the whole nucleus. All these possibilities in equation (1) will be divided into the mean field, short-range correlations and long-range correlations. Therefore, these three parts are related to each other, and they cannot be expressed by (0,0) due to the existence of the hidden-color states. In particular, the case that neither $(\lambda_{k},\mu_{k})$ is (0,0) and the intermediate coupling is also not (0,0) is the most common situation \cite{Bakker14,Abbas17}. Although the way of the division is still unclear, the result is clear. A simple case for all nuclei with $A\geq 2$ is that the nucleons in the hidden-color state are represented as (1,1), and their couplings are also (1,1), which makes this conclusion true.

Detailed studies should also consider other degrees of freedom of the nucleons, such as orbital, isospin and spin \cite{Bakker14}. Since this is not a specific calculation, there is no careful consideration. This Letter only considers about the color degree of freedom for the color confinement. In this Letter, I do not present a specific relationship between the hidden-color states and the SU(3) representation of the shapes of the nuclei, because this is very complex and may even be very difficult at the moment. I just focus on the simplest reasoning, and it is still very important.

One point to be emphasized here is that the color correlation with more nucleons must be weaker, and why such a hidden-color state associated with the entire nucleus appears. The possibility is that the more nucleons this state is related to, the bigger the number of such states will be. This requires further theoretical confirmation. In Ref. \cite{Bakker14}, it is shown that, the number of the hidden-color states with 9 quarks is much more than the one with 6 quarks \cite{Bakker14}. In a nucleus, the hidden-color states with various nucleon numbers may exist, but the most important may be the two-body correlation (the strongest), and the overall correlation caused by the hidden-color states with the nucleus number $A$ (the most possibilities). This is exactly what the nuclear experiments have been found out to be.

Since there are many hidden-color states in the nucleus, a possible result will occur. These hidden-color states will have quantum interference with each other, and if destructive interference occurs,  the long-range correlation may disappear and the magic number appears. If constructive interference occurs, a large deformation will occur. This is what to be focusing on in the future. If the magic numbers can be explained based on this result, it will further prove the correctness of this conclusion here.

Just as Bakker and Ji said:`` Nuclear systems are identified as color-singlet composites of quark and gluon fields, beginning with the six-quark Fock component of the deuteron. An immediate consequence is that nuclear states are a mixture of several color representations which cannot be described solely in terms of the conventional nucleon, meson, and isobar degrees of freedom: \emph{there must also exist hidden-color multi-quark wave-function components, i.e., nuclear states which are not separable at large distances into the usual color-singlet nucleon clusters}.'' The nucleons or their combinations may have the color-neutron component (0,0), but they can not the (0,0) representation. In this Letter, I give the first evidence that this hidden-color state should exist due to the existence of the Cd puzzle and its explanation with the SU3-IBM.

It should be emphasized that not one possibility in equation (1) corresponds to one division part, but the three parts are possible. It is not clear how to translate into the three parts, requiring supports from the experimental data. The influence of the color confinement in the nucleus is emphasized here, especially establishing the correlation between the low and high energy behaviors. Every possibility of the nuclear color confinement is to be considered, so all possibilities must affect the properties of the nucleus as a whole. If the hidden-color states exist, a \emph{global} effect must occur in the nucleus.

Note that the more number of nucleons involve in this hidden-color state, the associated energy may be lower (collective effect). When the number of nucleons is large enough, the energy of this hidden-color state may be the lowest. And that's exactly what the SU3-IBM expects and supports.

It is also important to emphasize that, although the color SU(3) symmetry is used in the discussion, the specific color state is not discussed here. The (0,0) represents a color-neutral state, but other $(\lambda_{k},\mu_{k})$ are just a phenomenological representation. This does mean that quarks or gluons are temporarily transferred from one nucleon to another, but the exact mechanism is not clear. Here, it is only emphasized that quarks and gluons cannot be fully bound within the nucleons in the nucleus. When approximately dividing into the mean field, short-rangle strong correlations and long-range correlations, although we still use the SU(3) symmetry, this may be an approximation and may have new physical significance, such as shape,  because this is a collective effect. If the mean field is $(\lambda_{m},\mu_{m})$, the short-range strong correlation is $(\lambda_{s},\mu_{s})$, the long-range correlation is $(\lambda_{l},\mu_{l})$, the division can be expressed as
\begin{equation}
(\lambda_{m},\mu_{m})\otimes (\lambda_{s},\mu_{s}) \otimes (\lambda_{l},\mu_{l}) =(0,0).
\end{equation}
Finding the division law and its limitations is at the heart of the further research.

First the mean filed is not (0,0). Although the magic number can be explained by the spherical single-particle potential, a potential field that does not deviate too much from the sphere is also possible. In the experiments, some researchers have also begun to find that some magic number nuclei exist a collective excitation \cite{Togashi18,Suchyta14,Taniuchi19,Tsunoda18}, so they are not spherical. Just as Garrett \emph{et al.} said:`` Rather than proceeding from spherical, closed-shell nuclei through a region of spherical vibrators before encountering deformation, deformation and shape coexistence may be confronted immediately, even at the \emph{closed shell}.'' \cite{Garrett19}  So the conclusion is credible. This result can also help us to find better single-particle potentials to understand the low-energy properties of nuclei.

\begin{figure}[tbh]
\includegraphics[scale=0.3]{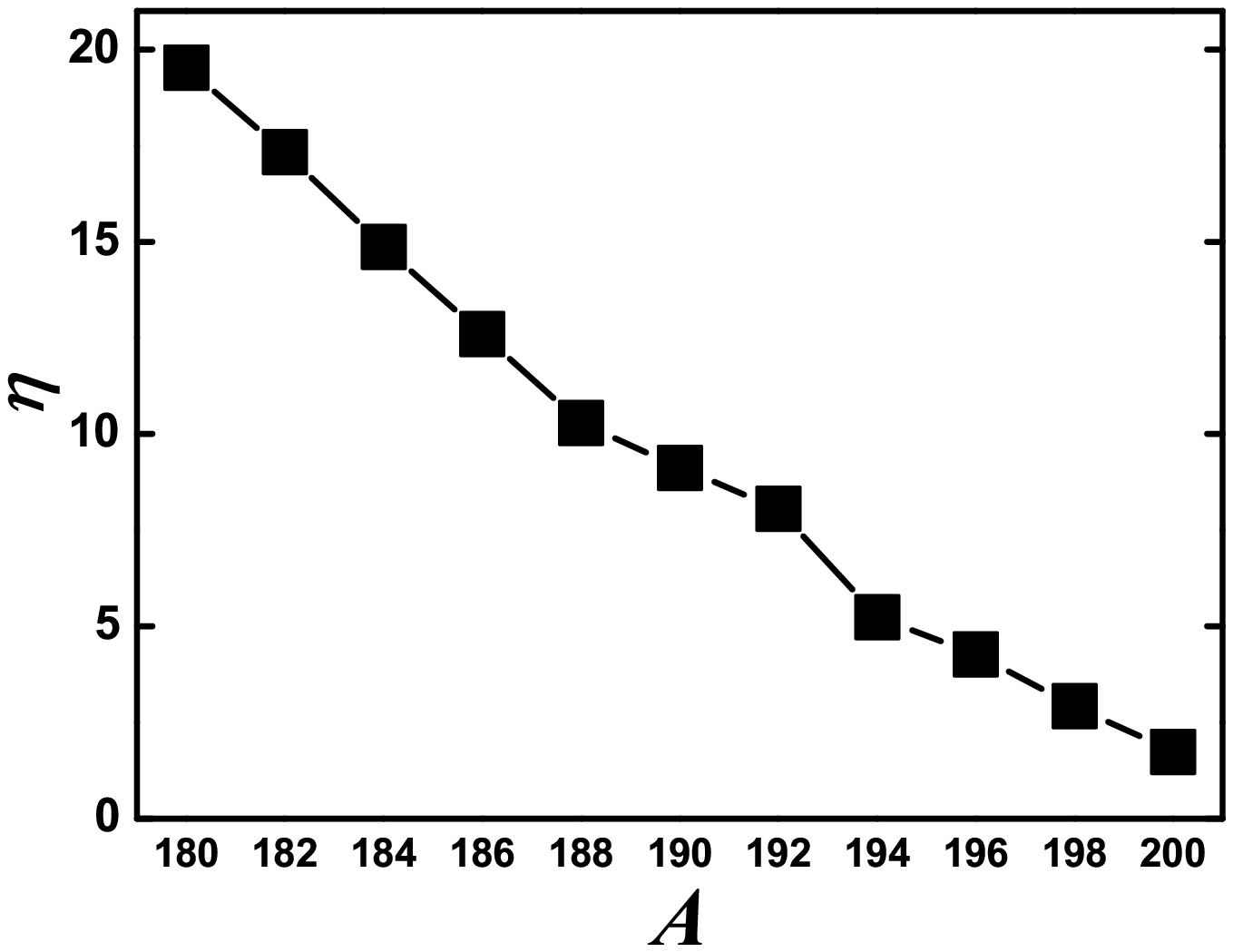}
\caption{The ratio of the two coefficients of the SU(3) second-order and third-order Casimir operators in Ref. \cite{wang23} with the number number $A$ in the Hf-Hg region.}
\end{figure}

Existing theories can provide more supports. It is found that the strong spin-orbit coupling disrupts the SU(3) symmetry, and only the collective states in the light nuclei can be classified by the SU(3) symmetry \cite{Elliott1,Elliott2,Harvey}. For the medium-heavy nuclei, this symmetry is broken \cite{Kota20}. However, the researchers found the pseudo-SU(3) symmetry \cite{Arima69,Hecht69,Draayer1,Draayer2}and the proxy-SU(3) symmetry \cite{Bonatsos17,Bonatsos23}. In particular, the latter can be used to discuss the medium-heavy nuclei, such as the shape transition from the prolate shape to the oblate shape. In this Letter, it can be found that this may not be by accident. This also supports the existence of the SU(3) symmetry from the perspective of the shell model.

Second the long-range correlations are not (0,0). This long-range correlations lead to the deformations of the nucleus. If they are not (0,0), it means that it cannot be spherical. This is exactly what the Cd puzzle is trying to reveal. This conclusion is also supported when the Cd puzzle is verified by the SU3-IBM. In the new spherical-like $\gamma$-soft mode, the SU(3) third-order Casimir operator is as important as the second-order one. This is a new kind of deformation. In Ref. \cite{wang23}, the shape phase transition from the prolate shape to the oblate shape in the Hf-Hg region is studied, and the ratio of the SU(3) second-order and third-order Casimir operators is shown in Fig. 1. This value decreases approximately linearly as the number of nucleons increases (the number of bosons decreases). Recently it is also found that the boson number odd-even effect \cite{Fortunato11,Zhang12} of the $^{196-204}$Hg really exists, in which the SU(3) third-order interaction plays the main role \cite{Wang}. This means that when the valence nucleons increase, the two-body residual interaction may not dominate. This is not quite the usual mean field viewpoint. I guess that the multiple possibilities of the color confinement in the nucleus in equation (1) let the low-energy two-body interactions interfere destructively, reducing it. The appearance of the magic numbers may arise from the complete quantum interference between the hidden-color states. For the Cd puzzle can be well explained by the SU3-IBM, the hidden-color state in equation (1) is much more important than expected.

These are the main conclusions to be given in this Letter. If the shape of the nucleus is not spherical and the SU3-IBM description is correct, this simple interpretation directly demonstrates that, in the nucleus, nucleons are colored, and the color degree of freedom plays an important role even in the low energy excitation in the nuclear structure. Although this idea has been speculated by some researchers, no evidence has been found in previous studies. In thie Letter I find the first clue.

\section{5. Discussion}

The deformation of the nucleus is a fundamental problem of nuclear structure. In the previous understanding, because the magic number nucleus is spherical, the deformation needs to occur when the number of valence nucleons is relatively large. The Cd puzzle \cite{Heyde16,Wood11,Garrett18} and the energy spectra of the magic nuclei \cite{Togashi18,Suchyta14,Taniuchi19,Tsunoda18} do not support such a conclusion. If the mean field is not spherical, then new collective excitation can emerge when the valence nucleons appear. If we can get the explicit expressions of $(\lambda_{m},\mu_{m})$ and $(\lambda_{l},\mu_{l})$, we will understand exactly what happens.

This hypothesis is strongly supported by the EMC effect, which needs to be further verified. The Cd puzzle is the direct consequence of this hypothesis. Although this explanation is easy, it fits with the results of the experiments. The conclusion that spherical nucleus does not exist can be confirmed by the Cd puzzle experiments and can be supported by other surrounding nuclei, such as Te, Pd nuclei \cite{Garrett18}. The results of these experiments can be well described by the SU3-IBM \cite{Wang22,WangPt,Wang24}, which further supports this hypothesis.

The SU3-IBM explains not only the normal states of the Cd nuclei, but also the B(E2) anomaly phenomenon \cite{Wang20,Zhang22,Zhang24,Teng24}, which cannot be explained by previous theories of nuclear structure \cite{166W,168Os,172Pt,170Os}. This model is also used to describe the prolate-oblate asymmetric shape evolution, which reveals the deformation of the prolate shape is nearly twice of the one of the oblate shape \cite{wang23}. This model can explain the properties of $^{196}$Pt at a better level \cite{Wang24,zhou24} and also the E(5)-like behavior of $^{82}$Kr \cite{zhou23}. All these mean that the SU3-IBM is able to give a more accurate description of the low energy excitations of the nuclei and is a more efficient model. So the conclusion that the spherical nucleus does not exist is plausible, which supports the correctness of the hypothesis.

The result is somewhat unexpected. For the first time, it explicitly points out a clear correlation between the low and high energy behaviors of the nucleus, and while the true reason for the correlation is still unknown, it does exist. So this hypothesis is supported by both experiment and theory. The role of the quarks and gluons in the nucleus, though not clear, exists and works at low energy excitation. This has been overlooked in previous studies. Of course, this is inevitable. Because the change of the quarks and gluons in the nucleus is the behaviors found in the high-energy region, it is difficult to expect that the EMC effect can affect the low-energy nuclear structure. This requires support from the experiments. The Cd puzzle, as well as its SU3-IBM description, becomes the possible evidence.

Since the EMC effect increases with the number of nucleons, it can be guessed that the effect of this color degree of freedom is more obvious for heavy nuclei. The fact is true. The SU3-IBM explains the prolate-oblate asymmetric phase transition in the Hf-Hg region, where the SU(3) third-order Casimir operator describes the oblate shape. The recent discovery of the boson number odd-even effect in $^{196-204}$Hg proves the correctness of this theory \cite{Wang}, and also indicates the emergence of previously completely unexpected phenomena in heavy nuclei.
These strange color-related phenomena may be very difficult to detect in light nuclei. It is not appropriate to draw conclusions from mere experiences with light nuclei. Thus the influence of the color confinement in the nucleus should be found in the nucleus with more nucleons, especially the medium-heavy nuclei.

How to understand the color confinement in the nucleus is important for the following studies. Equation (1) is the starting point of the discussions. The key thing is how to extract the mean field, short-range strong correlations and long-range correlations from all possibilities of the color confinement, the equation (1). This is an interesting and urgent question. It should be noticed that if this division rule exists, it must be sensitive to the nucleon number $A$. We expect that this may have something to do with the magic numbers and the reduction of the SU(3) second-order interaction. Quantum interference among the hidden-color states is a very attractive possibility.

Currently, the main tool for nuclear force researches is the effective field theory \cite{Machleidt24,Meissner09,Machleidt11,Kolck20}. In their discussions, the color degree of freedom is not considered, which means that the quarks and gluons are simply bound to the nucleons. The color confinement in the nucleus is not considered in these studies. It can be seen from this Letter that for a complete understanding of the nuclear force, even at low energy, quarks and gluons need to be considered. How to combine the idea of the effective field with equation (1) is an important work. Just as Miller said:``discovery of new phenomena is not well treated by scale separation techniques because new phenomena are often related to discovering a new relevant scale.'' ``Understanding the EMC effect involves understanding physics at all three length scales (the nuclear size, the internucleon separation distance and the nucleon size).'' ``all three scales are must be understood to truly understand the physics of nuclei.'' \cite{Miller20} If the division equation (4) exists, the idea of the effective field theory is still applicable, only adding a new limitation.

What needs to be emphasized is that although the reason for the appearance of the hidden-color states is still unclear, the low-energy collective behavior may be understood by the perturbative quantum chromodynamics \cite{Bakker14,Ji86}, emphasizing the exchange of the quarks and gluons among the nucleons. This is very important, if so, some of the conclusions given here can be proved. I look forward to further calculations.

The experimental study of the short distance interactions between nuclei is the key. I expect to find some experimental phenomena that are sensitive to the nucleon number $A$ in the nucleus \cite{Sargsian14} and correspond to the sensitivity of the shape of the nucleus to the nucleon number $A$. For the short distance interactions, the effect induced by the hidden-color states may be only a part, so it may not be easy to distinguish experimentally.

\section{6. Conclusion}

In conclusion, a simple explanation for the absence of the spherical nuclei is given based on the EMC effect, the short-range strong correlations, the Cd puzzle and other new experimental results, the calculation of the SU3-IBM and the hidden-color state. It can be obtained that the Cd puzzle is the result of the color confinement in the nucleus. If the hidden-color states exist, the spherical nucleus can not appear. It is clearly pointed out that the color degree of freedom is necessary even in low energy excitation of nuclear structure. The collectivity of the nucleus results from the hidden-color states. The correlations between the low-energy and high-energy behaviors in the nucleus does exist, and this is the first time to be found here, but what it is requires more further studies.

In this Letter a new theoretical framework to understand the nuclear forces, especially the short distance quark-gluon interactions, is established  phenomenologically and a possible big picture for the whole nuclear physics is drawn. Since this explanation is very simple and rigorous, the new connection deserves serious consideration. It does not conflict with previous effective field theories. The possibility of the color degree of freedom between the nucleons is considered here. The SU3-IBM supports this possibility and gives a guide for future theoretical developments. In future, the possible results of this new perspective will be discussed.

\end{document}